\documentclass{article}

\usepackage[nonatbib,final]{nips_2017}

\usepackage[utf8]{inputenc} 
\usepackage[T1]{fontenc}    
\usepackage{hyperref}       
\usepackage{url}            
\usepackage{booktabs}       
\usepackage{amsfonts}       
\usepackage{amsmath}       
\usepackage{nicefrac}       
\usepackage{microtype}      

\usepackage[square,numbers]{natbib}

\usepackage{graphicx}

\usepackage{subcaption}
\usepackage{multirow}
\usepackage{hhline}

\usepackage{amssymb}

\title{Smartphone-based paroxysmal atrial fibrillation monitoring with robust generalization}

\author{
	Tamas Madl \\
	Austrian Research Institute for Artificial Intelligence\\
	HeartShield Ltd.\\
	\texttt{tamas.madl@heartshield.net} \\
	 \And
	David Madl \\
	HeartShield Ltd.\\
	\texttt{david.madl@heartshield.net} \\
}

\begin{document}
	
	\maketitle

\begin{abstract}

Atrial fibrillation is increasingly prevalent, especially in the elderly, and challenging to detect due paroxysmal nature. Here, we propose novel computational methods based on heart beat intervals to facilitate rapid and robust discrimination between atrial fibrillation and sinus rhythm. We used low-cost Android smartphones, and recorded short, 30 second waveform data from 194 participants. In addition, we evaluated our approach on 8528 hand-held ECG recordings to show generalization.

Our approach achieves a sensitivity of 93\% and specificity of 94\% on 30 second waveforms, significantly outperforming previously proposed heart rate variability features and smartphone-based AFib detection methods, and substantiates the feasibility of real-world application on low-cost hardware.

\end{abstract}



\section{Introduction}
\label{intro}

Atrial fibrillation (AFib) can cause significant risks to health, such as thrombosis formation or cerebrovascular stroke. Estimated population prevalence ranges from 0.5 - 3\%, can reach 10 - 30\% in the elderly, and nearly one in six stroke events are caused by AFib \citep{reiffel2014atrial}. However, detecting paroxysmal atrial fibrillation remains challenging. A large percentage of all cases are unnoticed. 

Recently, smartphones have been suggested to facilitate home monitoring at any time, with simple handheld ECGs \citep{galloway2013iphone}, which the patient would need to purchase, or using the camera and flashlight as a substitute photoplethysmograph \citep{krivoshei2016smart}. These latter approaches obtain an approximate photoplethysmography signal from the brightness values of the built-in phone camera, and subsequently calculate inter-beat intervals from the temporal differences of maxima or other fiducial points. It has been argued that variability measures of beat intervals obtained from smartphones can be in good agreement with the same measures calculated from ECG \citep{peng2015extraction}, at least for higher-end devices with high quality built in camera sensors and circuitry.

However, low sensor quality, especially in lower-priced devices, as well as frequent motion artifacts and the lack of willingness or patience to sit completely still for extended periods of time pose serious challenges for both of these home screening methods. Evidence of reliable detection of AFib has so far been mostly limited to high end phones such as iPhones, to minimum measurement durations of two to five minutes (requiring very patient participants), and to a few dozen patients \citep{krivoshei2016smart}. 


Here, we describe a system with stronger generalization ability than previously proposed methods using robust measures of irregularity and multivariate analysis, and show its applicability in more practical settings; namely, up to 10x shorter measurements (30 s vs. 300 s), low-cost hardware, higher tolerance to movement artifacts, low false positive rate, and generalization across hardware types). 


\section{Methods}
\label{methods}


We constructed the model based on inter-beat intervals (IBIs) from Holter ECGs of 102 participants (see section `Training data`), and evaluated its performance on out of sample data from
\begin{itemize}
  \item{IBIs from smartphone recordings of 100 patients, recorded in a clinical setting}
  \item{IBIs from smartphone recordings of 200 participants, recorded in a home monitoring setting}
  \item{IBIs from 8528 hand-held ECG recordings, recorded in a home monitoring setting}
\end{itemize}

Signal processing and data analysis was performed using custom software written in Python. 
Each short signal was evenly resampled, detrended and bandpass filtered prior to beat detection. Signals were rescaled to zero-mean and unit variance to mitigate the extreme amplitude differences between different data sources (Holters, 12-lead ECGs and hand-held fingertip ECGs for home monitoring; smartphones of different models and makes). In the case of ECG signals, the QRS-detector of \citep{kim2016simple} was applied subsequently to extract inter-beat time intervals. 


ECG sampling rates varied between 128 Hz and 300 Hz, depending on data source (see details below). All smartphone PPGs were recorded by a custom developed app on Android smartphones, which, at the time of this publication, are limited by the operating system to optical sensor input with a maximum of 30 Hz sampling rate.

\subsection{Robust heart rate variability measures for AFib detection}

A large number of measures of heart rate variability (HRV) have been suggested in previous literature (see e.g. \citep{acharya2006heart} for a review), and the authors have developed an array of additional HRV measures, shown to perform well at discriminating other disease etiologies from healthy controls \citep{madl2016cinc}. For the present AFib detection model, we extracted the five best performing measures of HRV from this multitude of possibilities, applying a wrapper method of feature selection from machine learning literature \citep{guyon2003introduction}, yielding the following best-performing features in order of inclusion: 


\textbf{$F_1$. Standard deviation of the n'th order derivative of inter-beat intervals}, equivalent to the standard deviation of successive differences (of successive differences). In our experiments, n=5 led to the best results. The idea of taking the standard deviation of a derivative instead of the plain RR interval standard deviation (SDRR) has a long history in HRV literature - see e.g. \citep{brennan2001existing}.

Let $I= \langle i_1, ..., i_N \rangle $ denote the sequence of $N$ inter-beat intervals, such that each $i$ represents the temporal difference between two heart beats. Then, the first feature is defined as 

\begin{equation}
F_1(I) = \sqrt{\mathrm{Var	} \left[\frac{\mathrm{d}^5 I}{\mathrm{d}i^5} \right]}.
\end{equation}

\textbf{$F_2$. Histogram entropy}, a particular implementation of the popular entropy-based measures of heart rate variability (e.g. \citep{lake2002sample}), defined as follows. Let $h=\langle h_1, ..., h_B \rangle$ be a histogram of the frequency distribution of the inter-beat interval sequence $I$, discretized into $B$ bins, such that $h_j$ is the normalized frequency of all $i \in I$ that fall into the $j$th bin, and $w_j$ is the width of the $j$th bin. Then,

\begin{equation}
F_2(I) = -\sum_{j=1}^{B} h_j \log \left( \frac{h_j}{w_j} \right). 
\end{equation}

Since a larger number of bins increases susceptibility of noise and sensitivity to the particular bin edges, we chose the smallest viable number of bins ($B=2$), which yields the most robust generalization (more bin edges could allow more noisy beats to land in the `wrong' bin).

\textbf{$F_3$. Resemblance of the distribution of inter-beat intervals to a Rayleigh distribution}. Testing against the Rayleigh distribution is frequently used in physics to test for periodicity in time series of possibly regular events \citep{leahy1983searches}. After inclusion of the above two, more traditional HRV measures, the addition of this metric facilitated the highest increase in AFib predictive accuracy. We formalized the resemblance between the IBI distribution and a comparable (fitted) Rayleigh distribution as follows. Let $r(i; \sigma)$ denote the Rayleigh distribution:
of inter-beat intervals $i$, with a scale parameter $\sigma$. Let $\sigma_{ML}(I)$ be the maximum likelihood estimated scale parameter of the best-fitting Rayleigh distribution, given the series $I$. 


Furthermore, let the $\hat{f}(x)$ denote the kernel density estimate of sequence $x$, such that 

\begin{equation}
\hat {f}(x; \Sigma_s)={\frac {1}{n}}\sum _{j=1}^{n}\mathcal{N}(x-x_{j};0,\Sigma_s),
\end{equation}

$\mathcal{N}(.)$ is the Normal distribution, and the parameter $\Sigma_s$ is calculated using Silverman's rule of thumb bandwidth estimator \citep{silverman1986density}; that is, $\Sigma_s = \left(\frac{4 \sqrt{\mathrm{Var}[I]}^5}{3N}\right)^{0.2}$.

Then, $F_3$, the resemblance of the kernel density estimated distribution of inter-beat intervals $\hat{f}$ to the best-fitting Rayleigh distribution $r$ can be calculated numerically by summing up the absolute differences between the two functions:

\begin{equation}
F_3(I) = \sum_{j=1}^M \frac{|r(j \epsilon, \sigma_{ML}) - \hat{f}(j \epsilon, \Sigma_s)|}{\epsilon}.
\end{equation}

$\epsilon$ was chosen to be 1 ms (since none of the data sources hat a higher resolution / accuracy), and $M$ was chosen to cover most of the Rayleigh distribution. Specifically, M was ensured to be make the range of summation extend at least to the point where the Rayleigh distribution took on an amplitude smaller than $1\%$ of its peak, i.e., such that $r(M*\epsilon, \sigma_{ML}) < 0.01 r(\sigma_{ML}, \sigma_{ML})$.

\textbf{$F_4 \& F_5$. Measures of stochasticity based on a horizontal visibility graph of the IBI series}, in particular, the measures of \textbf{graph radius} and \textbf{disassortative entropy}, both shown before to be capable of outperforming previously proposed HRV measures in terms of predictive power for certain cardiovascular diseases \citep{madl2016cinc}. Structural properties of the original time series, such as periodicity or fractality, are preserved in a horizontal visibility graph representation; and it has been argued that they are well-suited for discriminating stochastic and chaotic processes \citep{luque2009horizontal}.


A horizontal visibility graph (HVG) $G_I$ can be constructed from the IBI series $I$ as follows. Let $t_j$ be the time that interval $I_j$ has `occurred'; that is, the time of the fiducial point in beat $j$ in the ECG or PPG signal. Then, $G$ is a network constructed of the series ${(t_1, i_1), ..., (t_N, i_N)}$, such that each IBI $i$ has a corresponding vertex, each pair of vertices corresponding to a pair of IBIs $i_a$ and $i_b$ is connected by an edge if both $i_a, i_b > i_N$ for all $a < N < b$ \citep{luque2009horizontal}. In other words, in a bar plot of all inter-beat intervals $I$, vertices corresponding to two particular IBIs are connected of an unbroken horizontal line can be drawn between them without intersecting any intermediate bars in the plot. 

Apart from their usefulness in differentiating noise from chaos arising from non-linear dynamics, HVGs are useful because they facilitate the application of any complex networks analysis tool to IBIs. According to preliminary feature selection, two particular complex network descriptors increased predictive accuracy the most: radius and disassortative entropy.

\textbf{$F_4$, HVG radius}, is defined as the minimum of eccentricities, where eccentricity $ecc(i_a)$ is simply the graph distance to the vertex $i_b$ farthest from it in the HVG: 

\begin{equation}
F_4(I) = min_{i \in V_{G_I}} \{ecc(i)\}
\end{equation}

\textbf{$F_5$, HVG disassortative entropy}, is a connectivity measure proposed by the author \citep{madl2016cinc}. It is defined as the entropy of the mixing matrix, measuring the information content of the tendency of vertices to connect to similar vertices. Whereas traditional assortativity \cite{newman2003mixing} is maximized by a graph always connecting high-degree vertices to other high-degree vertices, disassortative entropy is maximized by a graph in which the connectivity is randomized. Thus, it is a second-degree measure of stochasticity.

The disassortative entropy-based feature is defined as
\begin{equation}
F_5(I) = \sum_b \sum_a e_{ab} \log e_{ab},
\end{equation}

where $e_{ab}$ is the joint probability of the degrees of vertices $a$ and $b$ in HVG $G_I$ constructed from $I$. Numerically, $e_{ab}$ is the fraction of edges in $G_I$ that connect vertices with degree $a$ to vertices with degree $b$ (where degree in the graph theory sense simply means the number of edges incident to a vertex). See \cite{newman2003mixing}. 

Table \ref{sensspec} shows cross-validated accuracies using logistic regression, chosen to avoid overfitting the limited training data. Regularization and feature selection were based on a hold-out set of the training Holter ECGs, and were not adapted to smartphone PPG.


%
%


\section{Results}
\label{results}

The multivariate logistic regression model described above was constructed using solely the training dataset described above (research ECG datasets), and evaluated on out of sample testing datasets (smartphone PPGs or hand-held ECGs) not seen at training time. Table \ref{sensspec} shows a comparison to alternative physiological markers or biomarkers recently suggested for paroxysmal AFib diagnosis. On data from cheap smartphone hardware, our method displays significantly higher accuracy (94\% vs. 70\%), area under the ROC curve (AUC, 0.97 vs. 0.81), and specificity (94\% vs. 66\%), but slightly lower sensitivity (93\% vs. 95\%) compared to the recently proposed method by \citep{krivoshei2016smart}. A high specificity is important in a home monitoring setting for psychological, time, and financial reasons, which makes previously proposed methods difficult to apply to short-term, 30 second recordings obtained from low-cost smartphone hardware.

Table \ref{sensspec} also shows the performance of other physiological and biological markers recently proposed to diagnose paroxysmal atrial fibrillation in a clinical setting (see \citep{howlett2015diagnosing} for a recent review). Unlike the comparison with prior methods of smartphone-based screening, these markers were evaluated on separate patient cohorts, limiting the usefulness of direct comparison. Nevertheless, the order of magnitude of these performance indicators strongly suggests smartphone-based home monitoring to be a viable solution to detect unnoticed or asymptomatic paroxysmal atrial fibrillation, at very low cost to patients and the healthcare system.

\begin{table}[]
\centering
\caption{Sensitivities and specificities of smartphone home monitoring on 30 second heart beat interval sequences (first two columns), as well as physiological and biological markers (last three columns - separate cohorts) in detecting paroxysmal AFib on novel cohorts without retraining. Brackets show hand-held ECG results, numbers above concern smartphone PPG results.}
\label{sensspec}
\begin{tabular}{l||l|l||l|l|l}
 & \multicolumn{2}{c||}{ \begin{tabular}[c]{@{}l@{}}Smartphone\\ home monitoring\end{tabular} } & Resting ECG & Biomarker & Physiological \\ \cline{2-6}

            & \textbf{Ours}                                             & \begin{tabular}[c]{@{}l@{}}Krivoshei\\ et al., 2016\end{tabular}                                 & \begin{tabular}[c]{@{}l@{}}\footnotesize{P-wave dispersion}\\ \footnotesize{(Dilaveris 1998)}\end{tabular} & \begin{tabular}[c]{@{}l@{}}\footnotesize{NT-proBNP peptide}\\ \footnotesize{(Fonseca 2014)}\end{tabular} & \begin{tabular}[c]{@{}l@{}}\footnotesize{Left atrium size} \\ \footnotesize{+pump function} \\ \footnotesize{(Toh 2010)} \end{tabular} \\ \hhline{|=|=|=|=|=|=|}
Sensitivity & \begin{tabular}[c]{@{}l@{}}93\%\\ (90\%)\end{tabular}          & \begin{tabular}[c]{@{}l@{}}95\%\\ (96\%)\end{tabular} & 83\%                                                                                 & \textbf{100\%}                                                                     & 82\%                                                                                          \\ \hline
Specificity & \textbf{\begin{tabular}[c]{@{}l@{}}94\%\\ (76\%)\end{tabular}} & \begin{tabular}[c]{@{}l@{}}66\%\\ (62\%)\end{tabular} & 85\%                                                                                 & 70.4\%                                                                             & 91\%                                                                                          \\ \hline
ROC AUC & \textbf{\begin{tabular}[c]{@{}l@{}}0.97\\ (0.88)\end{tabular}} & \begin{tabular}[c]{@{}l@{}}0.81\\ (0.79)\end{tabular} & N/A                                                                                 & N/A                                                                             & N/A                                                                                          \\ \hline
Sample size & \begin{tabular}[c]{@{}l@{}}194 PPG\\ (5641 ECG)\end{tabular}           & \begin{tabular}[c]{@{}l@{}}194 PPG\\ (5641 ECG)\end{tabular}  & 100                                                                                  & 264                                                                                & 280                                                                                          
\end{tabular}
\end{table}
%

\section{Conclusion}
\label{conclusion}

Our results provide evidence that waveforms as short as 30 seconds are sufficient for robust detection, increasing utility for end-users who might be hard pressed to sit still without the slightest movement for the 5 minute periods used in previous work \citep{krivoshei2016smart}. We demonstrate high sensitivity and specificity on a much larger patient cohort than previous smartphone studies, with much lower hardware requirements. Finally, our approach seems to generalize across patient cohorts and even measurement devices. Although Table \ref{sensspec} shows differences between the smartphone PPG cohort and the hand-held ECG data, discrimination ability is high in both cases, suggestive of strong generalization. 

\bibliographystyle{elsarticle-harv}

\bibliography{sample}

\end{document}